\documentclass[12pt,showpacs,showkeys]{revtex4}
\usepackage{graphicx}
\usepackage{times}

\begin{document}

\title {CORRELATION-POLARIZATION EFFECTS IN ELECTRON/POSITRON SCATTERING FROM ACETYLENE:
A COMPARISON OF COMPUTATIONAL MODELS.}

\author{J. Franz$^a$, F.A. Gianturco$^b$\footnote{Corresponding author;
e.mail address: fa.gianturco@caspur.it. Fax: +39-06-4991.3305.},
K.L. Baluja$^c$,  J. Tennyson$^a$, R. Carey$^d$, R. Montuoro$^d$,
R.R. Lucchese$^d$, T. Stoecklin$^e$}
 \affiliation{$^a$ Department of Physics and
Astronomy, University College London, Gower Street, London WC1E 6BT,
United Kingdom\\ $^b$ Department of Chemistry, The University of
Rome La Sapienza and CNISM, Piazzale A. Moro 5, 00185 Rome, Italy\\
$^c$ Department of Physics and Astrophysics, University of Delhi, Delhi 110007, India\\
$^d$ Department of Chemistry, Texas A\&M University, College Station, Texas 77843-3255, USA\\
$^e$ Institut des Sciences Mol\'{e}culaires, CNRS-UMR 5255 , 351
Cours de la Lib\'{e}ration, F-33405 Talence, France}

\begin{abstract}
Different computational methods are employed to evaluate elastic
(rotationally summed) integral and differential cross sections for
low energy (below about 10 eV) positron scattering off gas-phase
C$_2$H$_2$ molecules. The computations are carried out at the static
and static-plus-polarization levels for describing the interaction
forces and the correlation-polarization contributions are found to
be an essential component for the correct description of low-energy
cross section behavior. The local model potentials derived from
density functional theory (DFT) and from the distributed positron
model (DPM) are found to produce very high-quality agreement with
existing measurements. On the other hand, the less satisfactory
agreement between the R-matrix (RM) results and measured data shows
the effects of the slow convergence rate of
configuration-interaction (CI) expansion methods with respect to the
size of the CI-expansion. To contrast the positron scattering
findings, results for electron-C$_2$H$_2$ integral and differential
cross sections, calculated with both a DFT model potential and the
R-matrix method, are compared and analysed around the shape
resonance energy region and found to produce better internal
agreement.
\end{abstract}

 \maketitle

\newpage

\section{Introduction}
The increase of interest in high-quality measurements involving
antimatter has attracted,in recent years,  the attention of
experimentalists and theoreticians in the field of molecular physics
\cite{CH,SG,SL,KSHI,SGB}. This has generated a wealth of new
information on the nanoscopic behavior of a broad variety of
molecular systems when they are made to interact with beams of
positrons at thermal and near-thermal energies. To understand the
interaction of positron beams with matter, it becomes important to
also be able to distinguish to what extent the additional features
of positron interaction with molecules (e.g. Ps formation and
positron annihilation) are related to positron dynamics and to
positron-electron correlation features. The study of even the
simplest of such observables, e.g. the elastic scattering integral
and differential cross sections occurring below the thresholds of
Positroniun (Ps) formation in polyatomic gases, already provides a
very useful testing ground for the theoretical and computational
models which are currently employed to analyze positron-matter
dynamics \cite{SG}.

While the electrostatic interaction can in principle be described
exactly by an essentially repulsive potential due to the molecular
network of (electrons+nuclei), different approximations for
correlation-polarization effects - the V$_{pcp}$ potential -  play
an essential role in deciding the quality of the adopted theoretical
model over the whole range of relevant distances between target and
the positron projectile.  As the projectile nears the target, in
fact, the repulsive Coulombic core further slows it down while the
attraction from the bound electrons increases and strongly modifies
its motion in the intermediate range of distances via a correlation
mechanism reminiscent of multiple scattering effects \cite{puska}.
This short-range effect should therefore be energy-dependent and
nonlocal and would asymptotically give rise to charge-induced
polarization effects, the leading term of which will be given via
the dipole-polarisability of the target molecule \cite{puska}. The
evaluation of the V$_{pcp}$ contribution to positron-molecule
interaction is therefore central to theoretical scattering
calculations and its correct evaluation within cross section
modelling studies is one of the stumbling blocks to the quantitative
interpretation of existing experimental findings.

In the present paper we therefore carry out a detailed comparison of
the results of different theoretical approximations for the
correlation-polarization forces by using the acetylene molecule as a
benchmark systen, in view of the availability of good quality
experimental data on this system and of its relatively simple
structure as a polyatomic target.

The structure of the paper shall be the following: in the next
Section II we will report in some detail an outline of the methods
employed to generate correlation-polarization and static potentials
while Section III will give and discuss our results for the elastic
(rotationally summed) integral cross sections. The differential
cross sections will be given by section IV and our conclusions will
be summarized in Section V.

\section{Scattering Equations and Interaction Forces}
In order to carry out our comparison between different treatments of
correlation-polarization forces, we have tested three different
approaches: we have employed the Single-Center-Expansion (SCE)
treatment of the scattering problem and included the V$_{pcp}$
potential in two different ways, i.e. the Density functional
modelling (DFT), used by us before \cite{occhigrossi}, and the
Distributed Positron model (DPM) also introduced earlier on
\cite{gibson1992} and employed within the SCE treatment. We have
then tested a Configuration Interaction (CI) procedure and
implemented it within the R-matrix (RM) treatment of the scattering
process \cite{BB}. Comparison with calculations by other groups
\cite{carvalho2000} will also be given for the sake of completeness.
In the following we shall provide a short outline of each of the
above computational methods.

\subsection{The SCE Scattering Equations}
In order to  obtain the scattering cross sections for polyatomic
molecules, we need to solve the Schrodinger equation for the total
system

\begin{equation}
(H - E)\Psi =0
\end{equation}

\noindent at the total energy E, for the corresponding wavetunction
$\Psi$. Here $H$ is the total Hamiltonian given by

\begin{equation}
H =\hat{H}_{mol}+\hat{K}+\hat{V}
\end{equation}

\noindent where $\hat{H}_{mol},\hat{K}$ and $\hat{V}$ represent the
operators of the molecular Hamiltonian, kinetic energy of the
scattered positron and interaction between the incident positron and
the target molecule, respectively. The $\hat{H}_{mol}$ further
consists, in general, of the rotational and vibrational parts

\begin{equation}
H_{mol}=H_{rot}+H_{vib}
\end{equation}

\noindent whereby we exclude, at the collision energies considered,
both electronic excitations and the Ps formation channels.

The total wavetunction $\Psi$ described in the molecular frame (MF)
reference system in which the z axis is taken along the direction of
the main molecular axis, is expanded around a single-centre (SCE) as

\begin{equation}
\Psi(\mathbf{r}_1...\mathbf{r}_Z,\mathbf{r}_p|\mathbf{R})=
\Psi_{mol}(\mathbf{r}_1...\mathbf{r}_Z|\mathbf{R})\varphi(\mathbf{r}_p|\mathbf{R})
\end{equation}

\noindent where

\begin{equation}
\varphi(\mathbf{r_p}|\mathbf{R})= \sum_{l\pi\mu
h}r_p^{-1}u_{lh}^{\pi\mu}(r_p|\mathbf{R})X_{hl}^{\pi\mu}(\mathbf{\hat{r}}_p)
\end{equation}

In equation (4), $\mathbf{r_i}$ represents the position vector of
the \emph{i}th electron among the Z bound electrons of the target,
taken from the center of mass. The quantity $\Psi_{mol}$ represents
the electronic wavefunction for the molecular target at the nuclear
geometry $\mathbf{R}$. The continuum function
$\varphi(\mathbf{r_p}|\mathbf{R})$ refers to the wavefunction of the
scattered positron under the full action of the field created by the
molecular electrons and by their response to the impinging positron
as described below. Each $u_{lh}^{\pi\mu}$ is the radial part of the
wavetunction for the incident particle and the $X_{hl}^{\pi\mu}$ are
the symmetry-adapted angular basis functions discussed earlier
\cite{GJ} which we will not repeat further here.

The label $\pi$ stands for the irreducible representation (IR),
$\mu$ distinguishes the components of the basis for each IR and
$\{\pi l\}$, respectively.

Since the molecular rotations and vibrations are often slow when
compared with the velocity of the impinging positrons considered in
the present study, we may apply the fixed-nuclei (FN) approximation
\cite{changFano} that ignores the molecular term of $H_{mol}$ in
equation (2) and fixes the values of all $\mathbf{R}$ at their
equilibrium locations in each of the target molecules. We then solve
the Schr\"{o}dinger equation in the FN approximation, make use of
the MF system rather than the space-frame (SF) reference system: the
two systems are related through a frame transformation scheme given,
for example, by \cite{changFano}.

After substituting equation (4) into (1) under the FN approximation,
we obtain a set of coupled differential equations for $u_{lv}$
where, for simplicity, $v$ represents $(\pi\mu h)$ collectively:

\begin{equation}
\left\{\frac{d^2}{dr_p^2}-\frac{l(l+1)}{r_p^2}+k^2\right\}u_{lv}(r_p|\mathbf{R})=
2\sum_{l'v'}\langle lv|\mathbf{V}|l'v'\rangle
u_{l'v'}(r_p|\mathbf{R})
\end{equation}

\noindent with

\begin{equation}
\langle lv|\mathbf{V}|l'v'\rangle =\int
d\hat{r}_pX_{lv}(\hat{r}_p)\ast V(r_p|\mathbf{R})X_{l'v'}(\hat{r}_p)
\end{equation}

Solving equation (6) under the boundary conditions that the
asymptotic form of $u_{l'v'}^{lv}$ is represented by a sum
containing the incident plane wave of the projectile and the
outgoing spherical wave we obtain the corresponding S-matrix
elements, $S_{l'v'}^{lv}$. The actual numerical procedure we have
employed to solve that equation was given in detail in
\cite{fag2000,fag1991}.

After transforming the MF quantities into the SF frame, the integral
cross section (ICS) for the  elastic scattering, rotationally summed
over molecular rotations, is given by

\begin{equation}
Q=\frac{\pi}{k^2}\sum_{lv}\sum_{l'v'}|T_{l'v'}^{lv}|^2
\end{equation}

\noindent where $T_{l'v'}^{lv}=\delta
_{ll'}\delta_{vv'}-S_{l'v'}^{lv}$

\subsection{The DFT Modelling of Correlation-Polarization}
The present treatment of the short-range part of the full V$_{pcp}$
interaction was first applied to positron scattering problems by
some of the authors \cite{occhigrossi} and is based on constructing
the correlation energy $\varepsilon^{e-p}$ of a localized positron
in an electron gas, and in further connecting it with the correct
asymptotic form of the spherical dipole polarizability component
contained within the full potential reported on the rhs of equation
(6). The quantity $\varepsilon^{e-p}$ had been originally derived by
Arponen and Pajanne \cite{arponen} by assuming that the incoming
positron can be treated as a charged impurity at a fixed distance
$r_p$ in an homogeneous electron gas, which is in turn described as
a set of interacting bosons representing the collective excitations
within the random phase approximation. Based on their work, Boronski
and Nieminen \cite{boronski} have given the interpolation formulae
of $\varepsilon^{e-p}$ over the entire range of the density
parameter $r_s$, which satisfies the relationship $\frac{4}{3}\pi
r_s^3\rho (\mathbf{r})$ = 1, with $\rho (\mathbf{r})$ being the
density of the Z bound electrons.

The relationship between the correlation potential V$_{corr}$ and
$\varepsilon^{e-p}$, which is consistent with the local density
approximation and a variational principle for a total collision
system with the size of the target, is given by the functional
derivative of $\varepsilon^{e-p}$ with respect to the electron
density in the target:

\begin{equation}
V_{corr}(\mathbf{r}_p|\mathbf{R})=\frac{\delta}{\delta\rho}\{\varepsilon^{e-p}[\rho(\mathbf{r}_e|\mathbf{R})]\}
\end{equation}

\noindent where in our treatment $\rho(\mathbf{r}_e|\mathbf{R})$
denotes the undistorted electronic density of the target, obtained
from accurate Hartree-Fock (HF) calculations, as a a function of the
molecular geometry and the r$_e$ coordinates of the bound electrons.
This quantity provides the probability of finding any of the
molecular electrons near the impinging positron once an analytic
ansatz is provided for $\varepsilon^{e-p}$ \cite{boronski}. Thus,
the total V$_{pcp}$ potential for the e$^+$-molecule system can be
assembled by writing, for each molecular geometry.

\begin{eqnarray}
&&V_{pcp}(\mathbf{r}_p|\mathbf{R})= \left\{
\begin{array}{l}
V_{corr}(\mathbf{r}_p|R)\;\; for\;\; \mathbf{r}_p<r_c\\
V_{pol}(\mathbf{r}_p|R)\;\; for\;\; \mathbf{r}_p>r_c
\end{array}\right.
\end{eqnarray}

The V$_{corr}$ is connected to the spherical part of the asymptotic
form of the polarisation at the position of $r_c$, usually around
4.0 a$_u$ for our systems. It corresponds to the outer crossing
between the potential contributions of eq.(10).

The total interaction potential $V_{tot}$ is therefore given by the
exact static interaction $V_{st}$ between the impinging positron and
the components (electrons and nuclei) of the molecular target , (for
its detailed form see, for example, \cite{occhigrossi}) plus the
$V_{pcp}$ given by equation \cite{carvalho2000} for each choice of
fixed molecular geometry $\mathbf{R}$:

\begin{equation}
V_{tot}(\mathbf{r}_p|\mathbf{R})=V_{st}(\mathbf{r}_p|\mathbf{R})+V_{pcp}(\mathbf{r}_p|\mathbf{R})
\end{equation}

\subsection{The Distributed Positron Model}
An alternative model correlation-polarization potential we have used
to describe the $V_{pcp}$ in Eq. (10) is the distributed positron
model (DPM) potential, $V_{pcp}^{DPM}$ \cite{gibson1990,fag2001}.
The form adopted here for the correlation-polarization
$V_{pcp}^{DPM}$ is based on a modification of the adiabatic
polarization approximation, which provides a variational estimate of
the polarization potential. In the adiabatic polarization
approximation, in fact, the positron is treated as an additional
"nucleus" (a point charge of +1) fixed at location $r_p$ with
respect to the center of mass of the atomic or molecular target. The
target orbitals are allowed to relax  in the presence of this fixed
additional charge and the decrease in energy due to the distortion
is recorded. The difference between the final and initial values of
the energy defines the adiabatic polarization potential at one point
in space. In order to describe carefully the spatial dependence of
this interaction, the calculations need to be performed on a rather
large number of three dimensional grid points.

However, due to nonadiabatic and short-range correlation effects,
e.g., virtual Ps formation, the adiabatic approximation
mayoverestimate the strength of the polarization potential for
smaller values of $r_p$, where the positron has penetrated the
target electronic cloud. The present model corrects for this by
treating the positron as a "smeared out" charge distribution rather
than as a point charge. If the scattering particle really were an
additional point charge, then the dominant short-range correlation
effect would be virtual hydrogen atom formation into ground and
excited states, and the $\delta$-function distribution of positive
charge at the center of mass would be correct. But, for a Ps atom,
the positive charge is not localized at the center of mass, and to
mimic this effect in computing the polarization potential we
represent the positron as a spherically symmetric distribution of
charge. This leads to a polarization potential that more closely
reflects the correct physics and that smoothly reduces to the
expected result for larger values of $r_p$ without any need to
select a crossover distance as done in the previous treatment.

Within the above model, one can, in principle, choose any reasonable
distribution that approximates the positive charge for virtual Ps
embedded in the near-target environment. Studies by Gibson
\cite{gibson1990} have shown that constructing the positron charge
distribution from the 1s STO-3G basis function with the tighter
Slater exponent of $\xi$=1.24 as recommended
\cite{fag2001,fag2003,fag2005,szabo} for a molecular environment
leads to accurate results. Once the $V_{pcp}^{DPM}$ potential is
calculated, it is combined with the static potential to yield the
total local interaction potential of Eq.(10).

After the development of the DPM to account for nonadiabatic
polarization effects in positron-molecule scattering, a somewhat
similar scheme was proposed by Bouferguene et al \cite{bouferguene}
for low-energy electron-H$_2$ collisions in which the polarization
interaction is computed by replacing the impinging electron with a
spherical Gaussian distribution of charge -1. Like the DPM for
positron scattering, this has the effect of reducing the
overestimation of the adiabatic potential near the target and can be
very efficiently implemented within a quantum chemistry framework.
However, for electron scattering these authors \cite{bouferguene}
found it necessary to use a distribution that varies with the
distance of the scattering electron from the molecular center of
mass and involves a semiempirical parameter.

In contrast, we have been able to obtain good agreement with various
measured positron-molecule scattering results within the DPM using
essentially the same procedure (i.e., without needing to adjust for
details of the target molecule) on a variety of systems
\cite{gibson1990,fag2001,fag2003,fag2005} as diverse as H$_2$ and
SF$_6$. This past success is the chief motivation for including the
DPM in the current study. In conclusion we construct the charge
distribution from a 1s STO-3G basis function with Slater exponent
$\xi$= 1.24 and once the $V_{pcp}^{DPM}$ potential is calculated, it
is combined with the static potential to yield the total local
interaction potential of Eq. (10), as mentioned before \cite{lane}.

\subsection{The ab-initio R-matrix approach for e$^+$/e$^-$ scattering}

\def\ra{{\bf r}_1}
\def\rb{{\bf r}_2}

\def\xb{{\bf x}_1}
\def\xN{{\bf x}_N}
\def\xp{{\bf x}_p}

\def\antisym{{\hat{ \mathcal{A}}}}

In the R-matrix method  the space is divided into two regions: an
inner region, defined by a sphere typically of radius 10 to 15
a$_0$, and an outer region. In the inner region the complicated
many-particle problem with correlation and polarization effects has
to be solved. In the outer region the target is represented by a
multipole expansion and the one-particle problem is solved by
propagating the R-matrix outwards. The R-matrix provides the link
between the two regions.

For molecular targets, the calculations within the inner region
reduce to a modified electronic structure calculation and standard
quantum chemistry codes have been adapted to this purpose. In
particular the UK molecular R-matrix codes \cite{jt225} use adapted
versions \cite{nob82a,jt204} of the Alchemy and Sweden-Molecule
codes for diatomic and polyatomic targets respectively. All
calculations discussed here employ updated versions of these codes.
For diatomic targets an implementation for positron scattering was
made some time ago \cite{jt50,jt86} and has recently been extended to
polyatomic targets \cite{jt411}.

The scattering wavefunction
for a given energy $E$ is built up as a linear combination \cite{jt225}

\begin{eqnarray}
\Psi(E) &=& \sum_k  A_{K}(E)  \Psi_{K} \;
\end{eqnarray}
where the coefficients $A_{K}$ are obtained by propagating the
R-matrix into the outer region. The R-matrix basis functions are
represented by the close-coupling expansion \cite{BB}
\begin{eqnarray}
\Psi_{K} &=&
\sum_A \sum_{i}  b_{Ai}^{K}
\antisym \left( \Xi^{Ne}_A \eta_{i} \right)
+ \sum_B \sum_{r} c_{B\,r}^{K} \Phi_{B\,r}^{N+1}
 \; ,
\end{eqnarray}
where the first sum runs over all products of target wavefunctions
$\Xi^{Ne}_A$ and electronic continuum orbitals
$\eta_{\overline{x}}$, and $\antisym$ antisymmetrizes all electrons.
The second sum runs over square-integrable functions
$\Phi_{B\,r}^{Z+1}$ of ($Z+1$) particles. In the following we
shortly describe the three different types of basis functions. The
target wavefunction $\Xi_{A}^{Ne}$ are obtained by diagonalizing the
Hamiltonian for the target molecule containing $N$ electrons. In
general the eigenfunctions of the target hamiltonian are linear
combinations of Slater determinants $\Lambda_{D}^{Ne}$. In the
scattering calculations the target functions are multiplied by the
continuum orbitals $\eta_{i}({\bf r}_{Z+1})$ which are occupied by
the scattered particle
\begin{eqnarray}
\antisym \left( \Xi_{A}^{Ne}  \eta_{i} \right) &=&  \sum_{D}
d_{D}^{A} \antisym \left( \Lambda_{D}^{Ne}({\bf r}_1 , \cdots, {\bf
r}_Z) \times \eta_{i}({\bf r}_{Z+1}) \right) \;.
\end{eqnarray}
Here the antisymmetrizer $\antisym$ acts only on the electrons.
Therefore in electron-molecule scattering calculations the
continuum orbital is anti-symmetrized with the target wavefunctions,
whereas in positron-molecule scattering calculations the extra
orbital is simply muliplied with the target wavefunction. In both
cases the coefficients $d_{D}^{A}$ are kept fixed (see e.g.
\cite{jt180} for an efficient algorithm). The square-integrable
functions are given by
\begin{eqnarray}
\Phi_{B\, r}^{Z+1} &=& \Phi_{B\,r}^{N+1}({\bf r}_1 , \cdots, {\bf
r}_Z , {\bf r}_{Z+1}) \\ \nonumber &=& \antisym \left(
\Lambda_{B}^{Ne}({\bf r}_1 , \cdots, {\bf r}_Z)
    \times \chi_{r}({\bf r}_{Z+1}) \right)\; .
\end{eqnarray}
where  $\Lambda_{B}^{Ne}$  is a $N$ electron function, like a Slater
determinant, and $\chi_{r}({\bf r}_{Z+1})$ is a square-integrable
spin-orbital. In the case of electron scattering the latter one is
anti-symmetrized with the former orbitals.

For electron scattering we have applied the static-exchange (SE) model and
the static-exchange-plus-polarization (SEP) model.
In both models the target is described by a Hartree-Fock wavefuncion.
In the SE model the additional electron can occupy all virtual orbitals
as well as
a set of  single centered
diffuse Gaussian orbitals which are used to represent
the continuum within the R-matrix sphere\cite{jt286}.
In the SEP model all single excitations are added to the configurations
generated in the SE model.

In the case of positron scattering we use the same spatial orbitals
for both electrons and positrons and define both a static and a
static-plus-polarization (SP) model in analogy with the SE and SEP
models employed for electron scattering. However, the positron can
occupy all orbitals, including those orbitals, which are occupied by
electrons. Furthermore, different spin-coupling rules apply in the
two cases since the positron spin is not coupled with the bound
electron spins.

\subsection{Semi-empirical R-matrix approach
using an enhancement-factor}

In order to model correlation effects which are not fully described in the
SP model, we have experimented with scaling the
electron-positron attraction integrals
by an empirically adjusted enhancement factor, $f$. These integrals
are the ones which are routinely referred to as two-electron
integrals in standard quantum chemistry language. Here we wish to only
increase the electron-positron attraction so do not alter the
corresponding electron-electron integrals.
\begin{eqnarray}
 (pq|\bar{r}\bar{s})_{\rm enh}
&=&  f (pq|\bar{r}\bar{s}) \\
 &=& f \int \phi_p({\bf r}_1) \phi_q({\bf r}_1)
 \left( - \frac{1}{|r_{1\overline{1}}|}  \right)
 \overline{\chi_r}({\bf r}_{\overline{1}}) \overline{\chi_s}({\bf r}_{\overline{1}})
 d{\bf r}_1 d{\bf r}_{\overline{1}}  \;\; .
\end{eqnarray}
Here $\phi_p({\bf r}_1)$ and $\phi_q({\bf r}_1)$ are electron orbitals,
$\overline{\chi_r}({\bf r}_{\overline{1}})$ and $\overline{\chi_s}({\bf r}_{\overline{1}}) $
are positron orbitals, and $|r_{1\overline{1}}| = |{\bf r}_{1} - {\bf r}_{\overline{1}}|$ is the
electron-positron distance.

This form of scaling can be justified as follows.
The second order contribution
of a M\o ller-Plesset type perturbative expansion
of the electron-positron correlation energy
\cite{tenno} is given by

\begin{eqnarray}
E^{(2)} =
\sum_{ia\overline{i}\overline{a}}
\frac{
(ia|\overline{i}\overline{a})^2}{
\epsilon_i - \epsilon_a +
\epsilon_{\overline{i}} - \epsilon_{\overline{a}}}
&=& \sum_{i\overline{i}}
(ii|\overline{i}\overline{i})^{(2)}  \;.
\end{eqnarray}

Here $\epsilon_i$ and  $\epsilon_a$ denote
energies of occupied and virtual electronic orbitals, respectively.
$\epsilon_{\overline{i}}$ and $\epsilon_{\overline{a}}$ are the
same for positrons.
Here we have introduced a second-order correction
$(ii|\overline{i}\overline{i})^{(2)}$
to the electron-positron attraction integral $(ii|\overline{i}\overline{i})$.
Since all denominators in the above expression are negative,
this correction is always negative, and
therefore has the same sign the integral
$(ii|\overline{i}\overline{i})$ itself.
By defining the pair-dependent enhancement factor

\begin{eqnarray}
f_{i\overline{i}} &=& 1 +
\frac{ (ii|\overline{i}\overline{i})^{(2)}}{(ii|\overline{i}\overline{i})} \; ,
\end{eqnarray}

\noindent the sum of first- and second-order contributions can be
re-expressed as

\begin{eqnarray}
E^{(1+2)} &=& \sum_{i\overline{i}} (ii|\overline{i}\overline{i}) +
\sum_{i\overline{i}} (ii|\overline{i}\overline{i})^{(2)} \\
\nonumber&=& \sum_{i\overline{i}} f_{i\overline{i}}
(ii|\overline{i}\overline{i}) \;.
\end{eqnarray}

If we assume, that the enhancement is the same for each
electron-positron pair, we can replace the pair-dependent
enhancement factor $f_{i\overline{i}}^{(2)}$  by an averaged
enhancement factor $f$

\begin{eqnarray}
E^{(1+2)}
&\approx&
\sum_{i\overline{i}}
f (ii|\overline{i}\overline{i}) \;,
\end{eqnarray}

\noindent as used in this paper. In our computations we have used
integrals of the more general type $(ij|\overline{i}\overline{j})$,
for which the second-order correction might be positive, resulting
in enhancement factors smaller than one. We have however used the
same enhancement factor for all integral types.

For large distances $r_p$ between the positron and the scattering
center the second-order contribution to the electron -- positron
correlation energy goes over to the asymptotic polarization
potential which is given by (in a.u.) \cite{itikawa2000}

\begin{equation}
-\frac{\alpha_0}{2 r_p^4} - P_2(\cos\theta) \frac{\alpha_2}{2 r_p^4}
\end{equation}

\noindent for a linear molecule. Here $\alpha_0$ ($\alpha_2$) is the
spherical (anisotropic) polarisability of the target molecule, and
$P_2(\cos\theta)$ is a Legendre polynomial, where $\theta$ is the
angle between the vector linking the positron to the molecular
centre-of-mass and that of the molecule. The long-range polarization
is included automatically, if not completely, in calculations which
use coupled-state expansions \cite{jt341,jt354} but not in the SP
model. Below we also discuss the influence of including asymptotic
polarization potential outside the R-matrix box, something that was
also tested in earlier R-matrix studies of positron -- molecule
collisions \cite{jt86,jt105}. In this paper the asymptotic
polarization potential has been neglected outside the R-matrix box.
This effect has been studied in more detail in \cite{fenh}. In all
calculations the quadrupole moment generated by the target molecule
was included in the outer-region.

The present approach is related, but not equivalent, to scaling the
positron charge or the electron charge. Such an approach, in fact,
would require a scaling of the positron -- nuclear repulsion
integrals or of the electron-nuclear attraction integrals in
addition to the scaling done here. Furthermore, a scaling of the
positron charge would require a scaling of the electrostatic
interaction outside the box, whereas a scaling of the charge of the
target electrons would introduce an attractive Coulomb-potential in
the outer region. No such scaling of charges was done here and
therefore no compensation of scaled charges is required.

To use $f=1$ products in the standard {\it ab initio} results of
previous R-matrix studies of positron -- molecule collisions. As
shown below, the results are very sensitive to the choice of $f$ and
values of $f$ only slightly bigger than unity yield surprisingly
good agreement with existing experiments.

\section{Results and Discussion}
\subsection{Low-energy positron scattering from C$_2$H$_2$ targets}
As mentioned in the Introduction, the present study intends to
select gaseous C$_2$H$_2$ as a benchmark system for an extended
comparison of methods. Some of the earlier work on this molecule,
both for positron and electron scattering studies, has been carried
out by some of the present authors \cite{fag2003b,fag2004,fag2006}.
Therefore the present work employs the same molecular geometry to
describe the equilibrium structure of the target, the same quality
of Hartree-Fock basis set and, whenever necessary, the same value of
the dipole polarisability coefficients for the long-range
polarization potential employed in the previous studies
\cite{fag2003b,fag2004,fag2006}.

The numerical convergence of the SCE has been carefully checked both
on the multipolar expansion of the potential (l$_{max}$=30) and on
the number of partial waves describing the scattering
e$^{+}$/e$^{-}$ particle.

In the R-matrix calculations for electron and positron scattering we
have used the DZP basis set of Dunning and coworkers, an R-matrix
box of a radius of 10 $a_0$, and the continuum basis set of Faure et
al.,  optimized for a box size of  12 $a_0$ \cite{jt286}. We have
taken the equilibrium geometry ($\rm r_{CC} = 1.208 \AA$ and $\rm
r_{CH} = 1.058 \AA$) optimized with the the Hartree-Fock method
using this basis set. For the positron-scattering calculations using
the enhancement factor, we have empirically found a factor $f =
1.004$ by comparing the calculating integral cross section with the
experimental results. Differential cross sections were calculated
using the program POLYDCS \cite{sg98}.

We report in figure 1 the elastic (rotationally summed) integral
cross sections for positron scattering, comparing various
theoretical results with experiments from Sueoka and Mori \cite{sueoka}
The earlier computations from Carvalho et al. \cite{carvalho2000} are also
given in the figure. The following considerations could be made from
a perusal of the results reported:
\begin{enumerate}
  \item The calculations at the static level, carried out with the
  same basis set and target description, provide the same results
  when using the SCE and the R-matrix treatment: both curves are
  given in the lower part of the figure and are essentially
  coincident, which is a good test for the two different scattering
  codes;
  \item When correlation-polarization forces are included, we see a
  dramatic change in the size and behavior of the cross sections
  obtained within the SCE approach and a remarkable agreement with
  the experiments down to very low collision energies;
  \item The use of the DFT and DPN approximations for V$_{pcp}$ potential
  yield results which are  in very good agreement with each other and in good
  accord with the experiments. The values of the DPM cross sections are consistently
  smaller than the DFT data especially around 1 eV and below. However, this
  difference keeps well below the experimental cross section in the considered energy range;
  \item The {\it ab initio} inclusion of polarisation effects within the
  R-matrix calculations using the SP model
  is seen, on the other hand, to be still far from
  convergence although it does give an upturn in the
  cross section dependance on collision energy near threshold. It is
  thus clear that the use of configuration interaction methods for treating correlation
  effects in positron scattering requires a much more extensive
  inclusion of additional states in order to be able to reproduce
  the experimental findings;
  \item The introduction of the empirical enhancement factor into the R-matrix method
  markedly improves the agreement with experiments of the calculated cross sections  at
  all energies.
  Compared with DFT and DPM calculations
  the RM cross sections are now
  slightly larger at low energies
  and slightly smaller at higher energies, although they display the correct energy dependence.
\end{enumerate}

If we now turn to the angular distributions from elastic scattering
processes, the results of figure 2 report the behavior from the
calculations using the DPM model within the SCE treatment of the
scattering problem. The two panels of figure report the differential
cross sections (DCS) over a broad range of collision energies; the
lower panel shows their values at higher energies from 4.0 to 10.0
eV, while the upper panel reports the same quantities down to 0.5
eV. The general trend of all angular distributions remains fairly
similar at all energies and indicates the presence of an intensity
"dip" that moves to lower angles as the energy increases. Thus, it
is only a shallow minimum in the DCS values beyond 130$^{\circ}$ at
0.5 eV while it moves down to 30$^{\circ}$ and becomes more marked
by the time the collision energy goes up to 10 eV.

The DCS calculated using the R-matrix method via the SP-model, with
and without enhancement factor, are given figure 3. The results
using the SP-model without the enhancement factor are given by thick
lines and do not show a strong forward peak. The calculations with
an enhancement fact, as in the DPM calculations, show a strong
forward peak. Indeed it would appear that at scattering energies
above 1 eV the smaller cross sections given by the RM calculations
are caused by their underestimation of the strong forward peak.

All computed cross sections turn out to be characterized by a large
forward peak and appear to converge to a similar limiting value for
all energies considered. In other words, the dominance of the
$l$=2 Legendre polynomial, which is associated with the dipole
polarizability coefficient, shows up in the angular shape of the DCS
at small angles and in the energy dependence of a sort of "magic"
angle in the angular distributions.

For the R-matrix calculations including the enhancement factor, the
value $f = 1.004$, has been used as a result of optimizing the
integral cross sections discussed before. The corresponding results
in figure 3 are given by thin lines: as mentioned before, the
inclusion of the enhancement factor introduces a much more marked
forward peaks at all energies. Furthermore these DCS are also
showing a minimum which moves to lower angles at increasing energy,
as found by the DPM results of figure 2. At energies above 4 eV,
however,  a second minimum appears, which is not observed in the DPM
calculations.

Differential cross section  for C$_2$H$_2$ have also been calculated
by Carvalho et al \cite{carvalho2003}, and measured by Kauppila et
al \cite{kauppila}. The angular distributions in both works are
showing strong forward peaks, in agreement with both the DPM
calculations of fig. 2 and the R-matrix calculations with the
enhancement factor in figure 3.

\subsection{Electron scattering from gaseous C$_2$H$_2$ targets}

As mentioned before, one of the objectives of the present work was
 to revisit the quantum observables associated with scattering
 electrons for processes involving the same target, the
C$_2$H$_2$ molecule, in order to compare the performances of
different as well as to investigate the differences with the
previous comparison carried out for positron-C$_2$H$_2$ scattering
attributes.

The results reported by figure 4 therefore present the calculations
at the exact Static+Exchange level carried out within the SCE
expansion but treating both the Static and the exchange potentials
as exact contributions since the latter was introduced as a
discrete, converged expansion over a separable, additional basis
set: for details see our discussion given in reference
\cite{FAG1994}. The dotted and dash-dotted curves reported by figure
4 clearly show the very good numerical agreement exhibited by the
SCE and RM methods at the SE level of modelling: thus, these data
further confirm the internal reliability of the two approaches in
describing e$^{+}$/e$^{-}$ scattering processes.

The inclusion of correlation-polarization forces, both within the
SCE-DFT and RM approaches, is reported by the dashed and solid
curves, respectively. Here again, the two treatments exhibit
reassuring agreement and similarities: the location of the
open-channel (shape) resonance feature falls between 2.2 and 2.5 eV,
the overall size of the cross sections is also rather similar over
the whole range of energies and both curves fall out asymptotically
to a very similar high-energy value. The threshold behaviour, on the
other hand, exhibits marked differences very close to zero energies,
although both methods would require further numerical tests within
that region before anything could be decided on the origin of the
discrepancies.

Another interesting set of comparisons is reported by the data shown
in the three panels of figure 5: we see there the behaviour of
computed and experimental differential cross sections as a function
of collision energies and for three different values of the
scattering angles. The top panel refers to $\vartheta$=40$^{\circ}$,
the middle panel to $\vartheta$=60$^{\circ}$ and the bottom one to
$\vartheta$=90$^{\circ}$, The solid lines refer to the R-matrix
calculations while the dashed curves report the SCE-DFT
calculations. The experimental data of Kochem et al. are from ref.
\cite{Kochem1985}.

We clearly see again that the treatment of correlation effects for
electron scattering processes are easier to model {\it ab initio} within
the R-matrix approach than their treatment for positron scattering
attributes: the RM results are fairly similar to the DFT
calculations without having to introduce any enhancement factor.
Furthermore, both computational models follow reasonably well the
experimental findings, especially for the case of the DFT results
while the RM results turn out to be slightly larger in the energy
region of the shape resonance. Both methods, however, reproduce well
the general energy dependence of the measured cross sections.

\section{Present Conclusions}
In the present work we have tried to analyze in some detail the
relative performances of different computational treatments for
obtaining low-energy scattering observables (integral cross sections
and angular distributions in the elastic channels) associated with
both positron and electron low-energy collisions with a specific
polyatomic target: the C$_2$H$_2$ molecule at its equilibrium
geometry.

In particular, we have first made sure that both methods are
comparable when the interaction forces are artificially simplified
to be given either by Static interaction only (for e$^{+}$
scattering) of by Static+Exchange interactions (for e$^{-}$
scattering): the present results clearly show that indeed both
sets of codes produce the same integral elastic (rotationally
summed) cross sections in spite of their implementing very different
computational procedures.

The next step has been to add within both treatments the effect of
correlation-polarisation forces as implemented within each code.

The results which compare our findings with experiments in the case
of the positron scattering show that the DFT/DPM modelling of
correlation forces, neither of which are entirely {\it ab initio}
are both able to yield remarkable agreement with observed
quantities, while the {\it ab initio} RM approach moves in the
correct direction but manages to reach the same good agreement only
when an empirical enhancement factor is included. Thus the RM
approach still requires further work  for positron scattering
studies on the treatment of correlation-polarisation effects in
order to achieve an acceptable level of convergence.

The corresponding analysis of computed angular distributions largely
confirms the above findings. The two methods produce very similar
results only when the empirical enhancement factor is employed
within the RM calculations.

Finally, the comparison between scattering attributes, as obtained
by the two methods analysed in the present work, has been extended
to electron collisions from the same molecular target. The results
reported by figures 4 and 5 indeed confirm the methods' reliability
at the SE level but also indicate the quality of the RM outcome when
electron projectiles are considered: both methods now yield similar
results  without using an enhancement factor within the RM
calculations. In general, however, the DFT angular distributions
appear to follow more closely the experimental data over a broad
range of energies.

In conclusion, the present benchmark calculations have allowed us to
attain the following, interesting results:
\begin{enumerate}
  \item that the existing implementations of the R-matrix and of the
  SCE-DFT approaches for treating e$^{+}$/e$^{-}$
  electronically elastic scattering off polyatomic
  targets provide essentially the same results when using the
  same interaction potentials;
  \item that the two codes model very differently the further
  inclusion of correlation-polarization forces and therefore show up
  at this level their differences of behaviour;
  \item that the DFT approach is found to be able
  to model such effects in ways which turn
  out to reproduce very well experimental results for positron and
  for electron scattering;
  \item that the RM approach shows faster CI convergence in the case
  of electron scattering data while still requires much larger
  expansion treatments in the case of positron scattering, as
  indicated by its empirical need of including an enhancement factor
  in the latter case.
\end{enumerate}

We have therefore shown that at least two different approaches to
multichannel scattering methods which provide observables for
polyatomic targets can be reliably employed in the future to yield
complementary information for the ever increasing range of
polyatomic gases that are being experimentally analyzed.

\section{acknowledgements}
JF, JT, and KLB acknowledge funding from both EPSRC and the Royal
Society through their India-UK exchange program. JF thanks the EIPAM
network of the European Science Foundation (ESF) for the award of a
fellowship at the beginning of the present work. FAG further thanks
the CASPUR Consortium for the availability of computational time for
the present project. This work was also supported by the Robert A.
Welch Foundation (Houston, TX) under grant A-1020. We also thank
T.L.Gibson and P. Nichols for providing the code PATMOL employed to
generate the DMP potential.

\begin{figure}[h!]
\begin{center}
\includegraphics[scale=0.50]{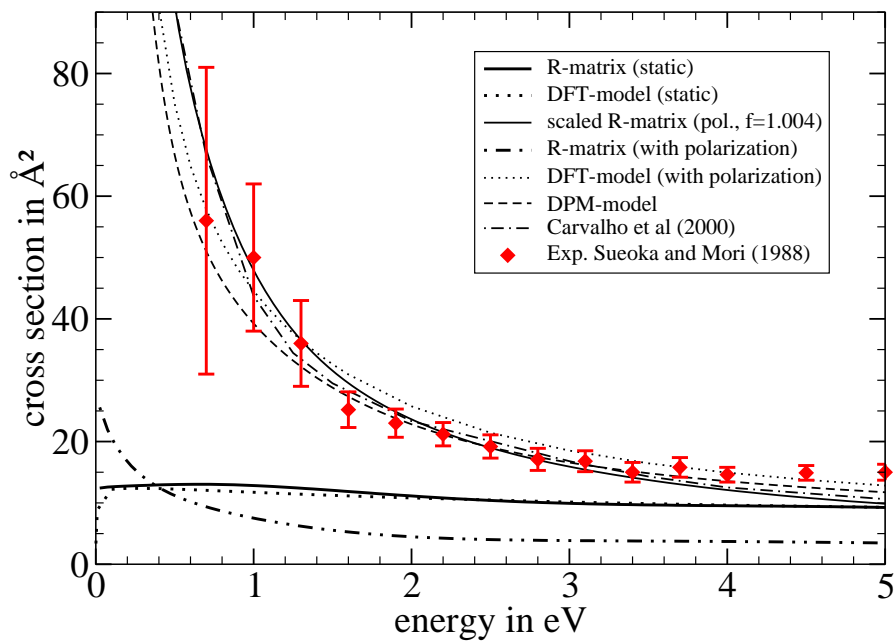}
\caption{Integral cross sections for positron-C$_2$H$_2$ scattering
using various theoretical methods. The experimental results of
Sueoka and Mori \cite{sueoka}, and the calculations of Carvalho et
al \cite{carvalho2000} are also given.} \label{figure1}
\end{center}
\end{figure}

\begin{figure}[h!]
\begin{center}
\includegraphics[scale=0.50]{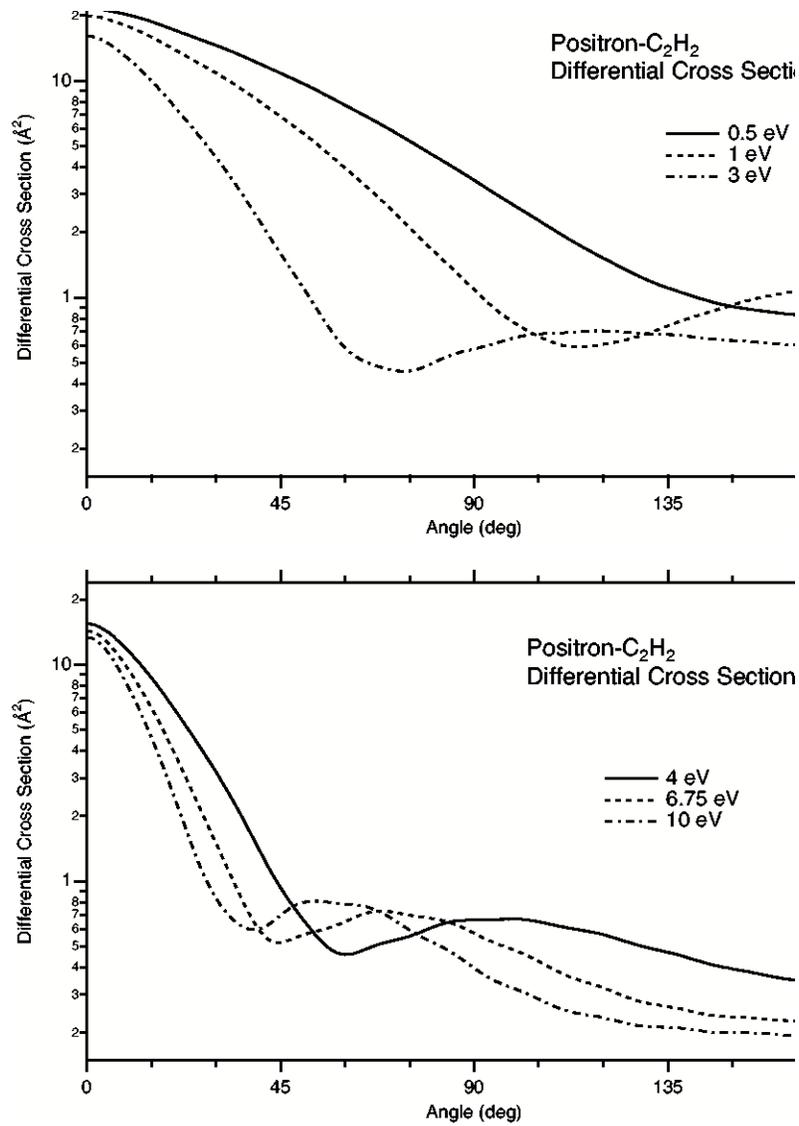}
\caption{Computed e$^+-C_2H_2$ elastic differential cross sections
using the DPM modelling of correlation forces. see text for
details.}\label{figure2}
\end{center}
\end{figure}

\begin{figure}[h!]
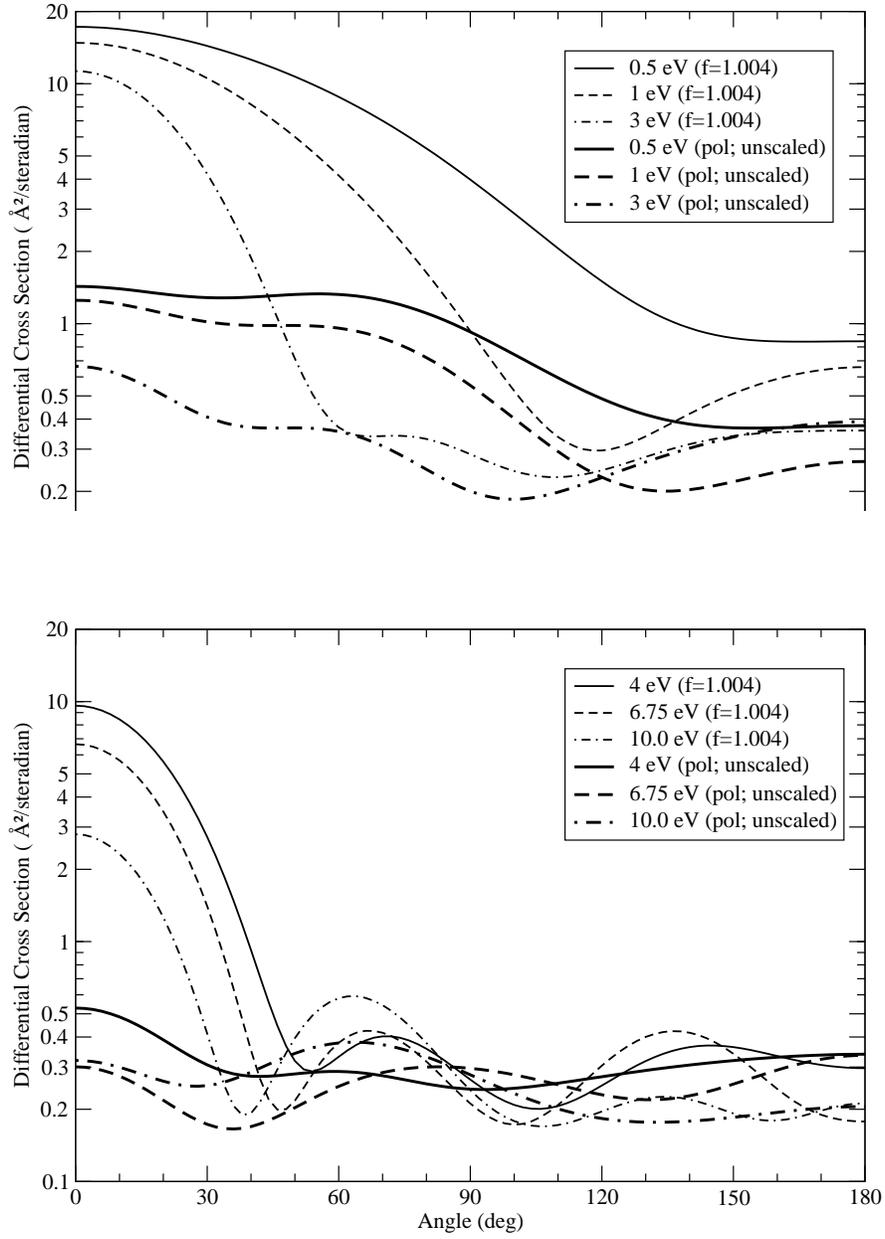

\begin{center}
\includegraphics[scale=0.50]{positron-dcs-lowEnergy.eps}\\
\includegraphics[scale=0.50]{positron-dcs-highEnergy.eps}
\caption{e$^+-C_2H_2$ elastic differential cross sections
calculated with the R-matrix method using the SP-model
with (thin lines) and without (thick lines) using
the enhancement factor ($f_{\rm enh} = 1.004$).}\label{figure3}
\end{center}
\end{figure}

\begin{figure}[h!]
\begin{center}
\includegraphics[scale=0.50,angle=0]{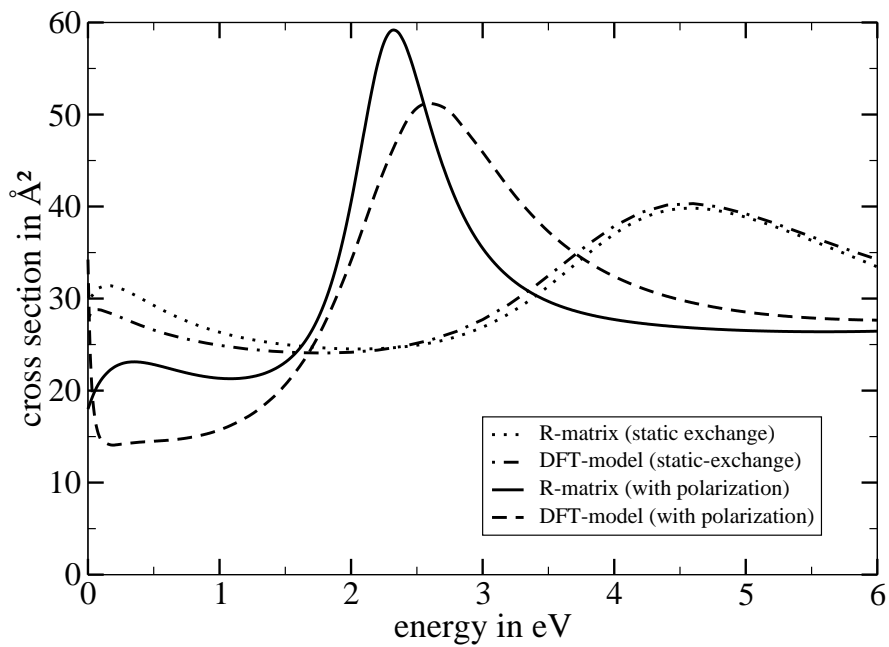}
\caption{Electron scattering: computed symmetry components of the
elastic integral cross sections (ICS) for electron scattering (upper
panel) and a comparison of total ICS obtained either using the DFT
correlation polarization model (solid line) or the static + exchange
method only (dashes)} \label{figure4}
\end{center}
\end{figure}
\newpage

\begin{figure}[h!]
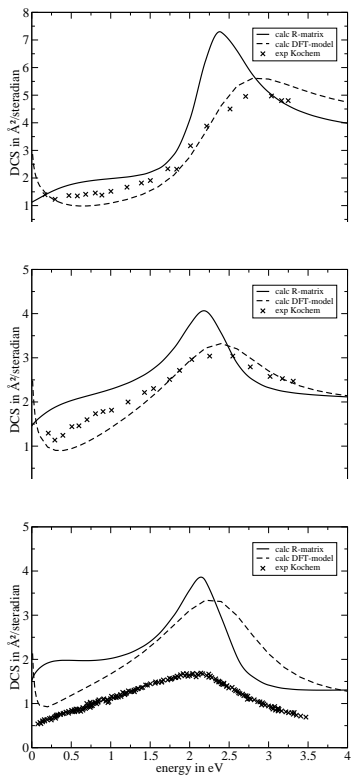

\begin{center}
\includegraphics[scale=0.20,angle=0]{electron-dcs-40.eps}\\
\includegraphics[scale=0.20,angle=0]{electron-dcs-60.eps}\\
\includegraphics[scale=0.20,angle=0]{electron-dcs-90.eps}
\caption{Computed $e^--C_2H_2$ angular distributions over the
location of the shape resonance energy and for three different
angular values ($\theta = 40,\; 60,\; {\rm and }\; 90$ degrees from
top to bottom panel). Experiments are from ref.\cite{Kochem1985}}
\label{figure5}
\end{center}
\end{figure}

\end{document}